%% file: main.tex
\documentclass[sigconf]{acmart}
\AtBeginDocument{%
  \providecommand\BibTeX{{%
    \normalfont B\kern-0.5em{\scshape i\kern-0.25em b}\kern-0.8em\TeX}}}

\copyrightyear{2024}
\acmYear{2024}
\setcopyright{rightsretained}
\acmConference[CHI '24 Getting Back Together Workshop]{CHI Conference on Human Factors in Computing Systems}{May 12, 2024}{Honolulu, HI, USA}

\usepackage{multirow}
\usepackage{framed,enumitem} 
\usepackage{graphicx}
\usepackage{xspace}
\usepackage{makecell}

\begin{document}

\title{Beyond Prompts: Learning from Human Communication for Enhanced AI Intent Alignment}

\author{Yoonsu Kim}
\email{yoonsu16@kaist.ac.kr}
\affiliation{%
  \institution{School of Computing, KAIST}
  \city{Daejeon}
  \country{Republic of Korea}
}

\author{Kihoon Son}
\email{kihoon.son@kaist.ac.kr}
\affiliation{%
  \institution{School of Computing, KAIST}
  \city{Daejeon}
  \country{Republic of Korea}
}
\author{Seoyoung Kim}
\email{youthskim@kaist.ac.kr}
\affiliation{%
  \institution{School of Computing, KAIST}
  \city{Daejeon}
  \country{Republic of Korea}
}

\author{Juho Kim}
\email{juhokim@kaist.ac.kr}
\affiliation{%
  \institution{School of Computing, KAIST}
  \city{Daejeon}
  \country{Republic of Korea}
}

\renewcommand{\shortauthors}{Yoonsu Kim et al.}




\input{sections/00_abstract}
\maketitle
\input{sections/01_introduction}
\input{sections/02_study-design}
\input{sections/03_study-result}
\input{sections/04_discussion}
\input{sections/05_conclusion}
\input{sections/06_ack}
\bibliographystyle{ACM-Reference-Format}
\bibliography{references}
\end{document}

%% file: sections/00_abstract.tex
\begin{abstract}
AI intent alignment, ensuring that AI produces outcomes as intended by users, is a critical challenge in human-AI interaction.
The emergence of generative AI, including LLMs, has intensified the significance of this problem, as interactions increasingly involve users specifying desired results for AI systems. 
In order to support better AI intent alignment, we aim to explore human strategies for intent specification in human-human communication. 
By studying and comparing human-human and human-LLM communication, we identify key strategies that can be applied to the design of AI systems that are more effective at understanding and aligning with user intent. 
This study aims to advance toward a human-centered AI system by bringing together human communication strategies for the design of AI systems.
\end{abstract}

%% file: sections/01_introduction.tex
\section{Introduction}
Recent advances in generative AI, particularly large language models (LLMs), have led to a paradigm shift in human-computer interaction. 
Instead of explicitly instructing machines how to perform tasks, users can now specify their desired outcomes in natural language, empowering a new paradigm of intent-based outcome specification ~\cite{AIFirstN88:online}.
By specifying desired outcomes in natural language, users can now leverage the capabilities of the AI to perform a wide range of tasks from creative content generation to problem-solving ~\cite{brown2020language, romera2023mathematical}.

While this shift has opened up exciting possibilities for various applications of AI, it has also introduced challenges in ensuring that AI systems reliably and correctly understand and align with user's intent ~\cite{christian2020alignment, Building32:online, terry2023ai}. 
Misalignment with the user's intent significantly impairs the user's experience when interacting with AI, making it difficult for them to achieve their goals. 
Specifically, a recent study based on user data found that the users most frequently experience dissatisfaction in terms of intent understanding when interacting with LLMs ~\cite{kim2023understanding}. 
Also, prior work of AI alignment reported that users have difficulties in knowing how to effectively ''prompt`` LLM and other generative AI systems to produce the desired outcomes ~\cite{10.1145/3591196.3593515, 10.1145/3491102.3501870, 10.1145/3544548.3581388}.

Users often struggle to deliver their intent clearly when interacting with generative models, whereas this issue is less pronounced in human-human interactions, like in conversation. 
Therefore, 
Humans have evolved sophisticated strategies for achieving intent alignment during conversation, such as active listening, clarification questions, and feedback ~\cite{norman2016we, Tomasello_Carpenter_Call_Behne_Moll_2005}.
We believe that these strategies can be adapted to the context of human-AI interaction.
By studying these strategies through a comparative study between human-human interaction and human-AI interaction, we aim to identify key principles adaptable to AI system design, enhancing intent alignment in human-AI interaction.
Specifically, we initially focus on verbal communication, such as text and voice, as this is the most common form of communication in human-AI interaction, and nonverbal cues can be difficult to interpret and may introduce additional complexity.

This paper presents a study that investigates human-human communication and extracts insights that can be applied to LLM-based interactions. We focus on the following research questions:
\begin{itemize}
    \item \textbf{RQ1.} What are the key verbal strategies that humans use to achieve intent alignment during conversation?
    \item \textbf{RQ2.} What are the benefits and limitations of these strategies?
    \item \textbf{RQ3.} How can these strategies be adapted and implemented in LLM-based user interfaces?
\end{itemize}
We believe that this study will provide valuable insights into how to improve intent alignment in human-AI interaction, and will ultimately lead to the development of more effective and user-friendly AI systems.

%% file: sections/02_study-design.tex
\section{Comparative study: Human-Human Interaction vs Human-AI Interaction}
\textbf{Study Design}
In order to investigate how human-human communication supports intent alignment compared with human-LLM interaction, we designed a comparative study. 
The study involves two participants in the following two roles: (1) \textbf{User}: the user is the primary role who interacts with the LLM and the human assistant to complete a given task, (2) \textbf{Assistant}: the assistant is a human participant who has the role of helping the user complete a given task using LLM. 
This study employs a within-subject design, where the \textbf{user} engages in two distinct conditions: \textbf{user}-LLM interaction (human-LLM interaction) and \textbf{user}-\textbf{assistant} interaction (human-human interaction).
In our study, we use GPT-4~\cite{openai2023gpt4} as an LLM model, and in the \textbf{user}-\textbf{assistant} interaction condition, the \textbf{assistant} can utilize the same LLM (GPT-4) to provide appropriate support to the \textbf{user}.

\smallskip
\noindent
\textbf{Task Selection}
Given the importance of observing intent alignment processes and the variability in user preferences and needs, selecting tasks that catered to diverse user needs was crucial.
To effectively observe the process of intent alignment, we intended to design a task that can motivate the users.
Therefore, we asked the study participants about tasks they personally found motivating beforehand.
Subsequently, we selected two types of tasks: ones that involve relatively straightforward criteria for defining desired outcomes (e.g., choosing gifts for family/friends) and tasks with more complex criteria (e.g., learning web development skills).

\smallskip
\noindent
\textbf{Study Procedure}
Eight participants (3 male, 5 female, average age 25.8) with prior experience using LLMs were recruited for this study. 
They were paired into four distinct teams, each consisting of a \textbf{user} and an \textbf{assistant}. 
Two teams were assigned to each task among the four teams.
During the study sessions, the \textbf{user} engaged in conversations with both an LLM and with a human \textbf{assistant} for the same task. 
When conversing with LLM, the \textbf{user} used the ChatGPT (model: GPT-4) interface. When conversing with human \textbf{assistant}, they had the option to use voice, and a shared document~\footnote{https://docs.google.com/} facilitated the conveyance of information that might be challenging to express verbally. 
The time limit for each task was set at a maximum of 15 minutes. 
After finishing two tasks, participants underwent semi-structured interviews to provide insights into their experiences.
The study sessions lasted approximately one hour and were conducted using the zoom~\footnote{https://zoom.us/} with cameras turned off to eliminate non-verbal cues.

%% file: sections/03_study-result.tex
\section{Study result: Intent Alignment Strategy}
\textbf{Study Result}
With participants' consent, we recorded and transcribed conversations and interviews from the study sessions and analyzed them to identify the strategies employed by individuals to achieve intent alignment.
While a thorough analysis is ongoing, we want to share our initial results about the strategies adopted by human assistants and the strategies exhibited by users during conversations with human assistants, which were not evident in interactions with LLMs.
\input{tables/assistant-strategy}
\input{tables/user-strategy}
The summarized results are presented in table ~\ref{tab:assistant} and ~\ref{tab:user}.

\smallskip
\noindent
\textbf{Assistant's Strategies}
Our initial findings indicate that human assistants better align with user intent by (1) actively requesting information from the user, (2) providing tailored responses based on previous conversation history and user responses, and (3) actively seeking feedback on responses and interactions from the user.
We found that human assistants proactively engage with users' inquiries, requesting additional information, asking clarification questions, or inquiring about potential areas for improvement in their responses.
It makes users provide more diverse and detailed contextual information, and it helps intent alignment.
Moreover, we observed that when the human assistant provides responses to the user, it filters and offers information gradually, reflecting on previous conversations with the user, rather than presenting a large amount of information at once.
For example, when the human assistant used LLM to respond to the user, they filtered and provided only the information they deemed necessary for the user based on previous conversations, amidst the many options provided by the LLM.
On the other hand, we observed passive behavior in LLM interactions, only responding passively to user prompts and failing to go further from user prompts.
Additionally, LLMs tend to present large lists at once in a response, often containing excessive, general, or even irrelevant information.
It causes users to perceive that their intentions are not fully reflected in the responses.

\smallskip
\noindent
\textbf{User's Strategies}
Beyond comparing the differences between human assistants and LLMs, we delved into analyzing interaction patterns when users conversed with LLMs compared to human assistants. 
Through this analysis, when the user interacts with the human assistant, we found that users (1) actively provided feedback in case of dissatisfaction with responses, (2) often interrupted the middle of the assistant's responses to ask questions if there were unknown parts, (3) actively offered additional context while responding to the assistant's questions, and (4) requested answers from the assistant's perspective.
If we examine the reasons why users exhibit these different behavioral strategies, it can be attributed to the differences in how human assistants and LLMs provide responses to users. 
Unlike human assistants, who provide a few filtered suggestions gradually, LLMs generate a large amount of suggestions at once. Therefore, users tend to behave by selecting their preferred options from the many suggestions provided by LLMs, rather than providing feedback on each suggestion to convey their intent more clearly. 

A significant factor contributing to users exhibiting different behavioral strategies may be the ability to converse using voice.
Unlike text, voice communication is conducive to providing feedback in short turns and displaying reactions while the other party is speaking. 
Users can interrupt in the middle if there are parts they do not understand and can also provide ambient reactions like ``uh-huh'' or ``okay'' intermittently if they understand the human assistant's answer.

Another reason may be due to the difference in the active/passive attitudes of human assistants and LLMs. Human assistants demonstrate an active attitude in conversations with users by actively asking questions or requesting feedback. Therefore, users tend to adopt a similar active communication style with human assistants or reveal their intent more clearly by responding to the assistant's questions.

%% file: tables/assistant-strategy.tex
\begin{table*}[ht]
\resizebox{\textwidth}{!}{
\def\arraystretch{1.1}
\begin{tabular}{l|l}
\toprule
\textbf{Type} & \textbf{Assistant's strategy} \\ \hline
\midrule
\multirow{6}{*}{Information Request to User} & - Requests information directly to user \\
 & - Asks questions to understand the user's context \\
 & - Asks questions to understand the user's context - Previous user's response specification question \\
 & - Asks why the user said something. \\
 & - Asks clarifying questions about the user's previous response. \\
 & - Asks for examples \\ \hline
\multirow{3}{*}{Reflecting Previous Conversation} &  - Maintains previous recommendations while incorporating the user's previous response. \\
 & - Provides answers reflecting the context of the previous conversation with the user. \\
 & - Provides information that the user may need based on the context of the conversation. \\ \hline
\multirow{5}{*}{Feedback Request to User} & - Asks for the user's opinion when providing a response. \\
 & - Asks which response the user liked best. \\
 & - Asks if there is anything that can be improved in the response. \\
 & - Asks if the user has any additional questions. \\
 & - Asks if the user is satisfied with the response. \\
\bottomrule
\end{tabular}
}
\caption{The types of strategies adopted by the \textbf{assistant} during the conversation with the user to achieve intent alignment.}
\label{tab:assistant}
\end{table*}

%% file: tables/user-strategy.tex
\begin{table*}[ht]
\scriptsize{
\resizebox{\textwidth}{!}{
\def\arraystretch{1.2}
\begin{tabular}{p{0.27\columnwidth}|p{0.65\columnwidth}|p{0.60\columnwidth}}
\toprule
\textbf{Type} &
  \textbf{User's behavior with human assistant} &
  \textbf{Comparison to User's behavior with LLM} \\ \hline
\midrule
\multirow{4}{*}{\makecell[l]{Feedback on Dissatisfactory \\Assistant Responses}} &
   - Evaluates the Assistant's response and provides the context/reason of the evaluation &
  \multirow{4}{*}{\makecell[l]{When talking to LLM, there are many options that LLM \\ suggests at once, so users tend to move on to the parts \\ they don't like and focus on the good parts of the answers \\ to continue the conversation.}} \\
 &
  - Informs problems in Assistant response &
   \\
 &
  - Corrects misunderstandings in Assistant response &
   \\
 &
  - Asks questions about unknown parts in Assistant response. &
   \\ \hline
Interrupting Assistant Mid-Response &
  - Interrupts in the middle of the Assistant's response and asks something &
  Users and LLM rarely interrupt during the conversation \\ \hline
\multirow{2}{*}{Provide Additional Context} &
  - Provides additional context when answering the Assistant's questions. &
  \multirow{2}{*}{\makecell[l]{LLM responds passively to users' prompts and does not \\ actively ask questions. Therefore, the user rarely gives \\ additional context while answering LLM's questions.}} \\
 &
  - Provides additional context on their own after the conversation. &
   \\ \hline
\multirow{2}{*}{\makecell[l]{Request an Assistant \\ Perspective Answer}} &
  - Asks for Assistant's opinion &
  \multirow{2}{*}{\makecell[l]{This could be because the user perceives the human assistant \\ as a person.}} \\
 &
  - Asks about Assistant's experiential content. &
   \\ \hline
\end{tabular}
}
}
\caption{The types of strategies exhibited by the user during the conversation with the human assistant to aid intent alignment, and comparison of user behaviors during interactions with LLMs.}
\label{tab:user}
\end{table*}

%% file: sections/04_discussion.tex
\section{Discussion}
Building upon our study findings, we aim to delve into a comprehensive discussion regarding the strategies humans employ to achieve intent alignment during human-human conversations, their respective advantages and limitations, and how these strategies can be integrated into human-AI interaction, specifically into LLM-based user interfaces.

Firstly, \textbf{incorporating proactive engagement features} within AI interfaces can enhance intent alignment. 
Recently, many systems based on LLMs are seeking to introduce proactive features. 
These systems primarily provide users with prompt suggestions(e.g., Notion~\cite{NotionAI53:online}, Grammarly~\cite{AIWritin58:online}). 
Subramonyam et al. also discussed that proactively suggesting ideas for prompting users is a useful design pattern that helps users obtain the intended answers~\cite{unknown}.
However, by analyzing the behavioral strategies of human assistants in our study, we found that, beyond suggesting prompts for users, actively seeking information from users and requesting feedback on responses contributes to intent alignment.
Therefore, designing AI interfaces that mimic this proactive behavior, such as prompting users for additional context or feedback, could improve alignment between user intents and AI responses.

Secondly, the findings highlighted \textbf{the importance of tailoring responses} to the user's context and preferences. 
Human assistants adjusted their responses based on previous conversations and the user's specific needs, leading to more aligned interactions. 
To achieve more aligned human-AI interaction with individual users' context and intent, AI interfaces should be designed to dynamically adapt responses based on user history.
We can get insights from recent studies~\cite{dombi2022common, 10.1145/3491102.3501972} that have designed natural language interfaces based on the Grice’s Cooperative principles ~\cite{grice1975logic} by considering the Recipient Design approach~\cite{unknown}.
However, it is necessary to discuss how such design approaches can be applied to interfaces for generative AI, including LLMs. 

Furthermore, \textbf{emulating the advantages of voice interfaces} in interactions with LLMs can facilitate intent alignment. 
Voice interfaces offer several benefits, including natural and fluid communication, real-time feedback, and short-turn conversations, allowing for quick exchanges of information.
While conversing with LLMs primarily through text, participants found it advantageous to use voice communication with human assistants. Incorporating flexible communication channels into AI interfaces can accommodate diverse user preferences and enable more effective communication of complex information.

%% file: sections/05_conclusion.tex
\section{Conclusion}
In conclusion, our study explores key strategies for enhancing intent alignment in human-AI interactions, compared with human-human interaction. We identified the intent alignment strategy that each user and human assistant takes in their conversation.
Human assistants demonstrate proactive engagement, actively seeking information and feedback and reflecting previous conversation with users.
While users adapt by offering feedback and interrupting for clarification.
Our discussion underscores the importance of incorporating these strategies into AI interface design. 
These discussions can serve as a foundation for designing more effective and user-friendly AI interfaces, addressing challenges in user intent understanding and alignment with LLMs.

%% file: sections/06_ack.tex
\begin{acks}
This work was supported by Institute of Information \& Communications Technology Planning \& Evaluation (IITP) grant funded by the Korea government(MSIT) (No.2019-0-00075, Artificial Intelligence Graduate School Program (KAIST)).
This work was also supported by Institute of Information \& Communications Technology Planning \& Evaluation (IITP) grant funded by the Korea government (MSIT) (No.2021-0-01347, Video Interaction Technologies Using Object-Oriented Video Modeling).
\end{acks} 